# Atmospheric aerosol diagnostics with UAV-based holographic imaging and computer vision*

Nathaniel R. Bristow, Nikolas Pardoe, and Jiarong Hong

*Abstract*—Emissions of particulate matter into the atmosphere are essential to characterize, in terms of properties such as particle size, morphology, and composition, to better understand impacts on public health and the climate. However, there is no currently available technology capable of measuring individual particles with such high detail over the extensive domains associated with events such as wildfires or volcanic eruptions. To solve this problem, we present an autonomous measurement system involving an unmanned aerial vehicle (UAV) coupled with a digital inline holographic microscope for *in situ* particle diagnostics. The flight control uses computer vision to localize and then trace the movements of particle-laden flows while sampling particles to determine their properties as they are transported away from their source. We demonstrate this system applied to measuring particulate matter in smoke plumes and discuss broader implications for this type of system in similar applications.

## I. INTRODUCTION

Numerous natural and anthropogenic processes, such as wildfires, volcanic eruptions, or pollution from smokestacks at power plants, result in the emission of harmful aerosols and particulate matter (PM) into the atmosphere. These emissions, and their subsequent dispersion, are associated with significant concerns for public health through air quality impacts [1, 2] as PM2.5 and PM10 particles can lodge deep into the lungs, as well as climate feedbacks through effects on radiative forcing [3, 4]. Modeling the dispersion of such particles and understanding the details of their impact on health and the environment, requires knowledge of particle characteristics such as size, morphology, and composition [5-7]. However, conventional particle diagnostics based on light scattering are insufficient to resolve such details [8-10]. Moreover, the small size of the aerosolized particles, coupled with the vast scale of motions associated with transporting atmospheric winds, imposes a challenge for measurement.

The challenge presented is to sample particles with high enough detail to be able to characterize individual particle properties while measuring across a large spatial domain. Remote sensing, such as with lidar, can measure across large domains but at the expense of particle-level detail [7]. There is, therefore, a need for *in situ* measurement approaches that can be paired with intelligent robotic platforms to efficiently and optimally sample aerosol data over large domains and with high resolution. Current measurement technologies are not capable of meeting these demands [8-12], and it is, therefore, the aim of the present work to implement an intelligent UAV-based particle measurement system to accomplish these objectives.

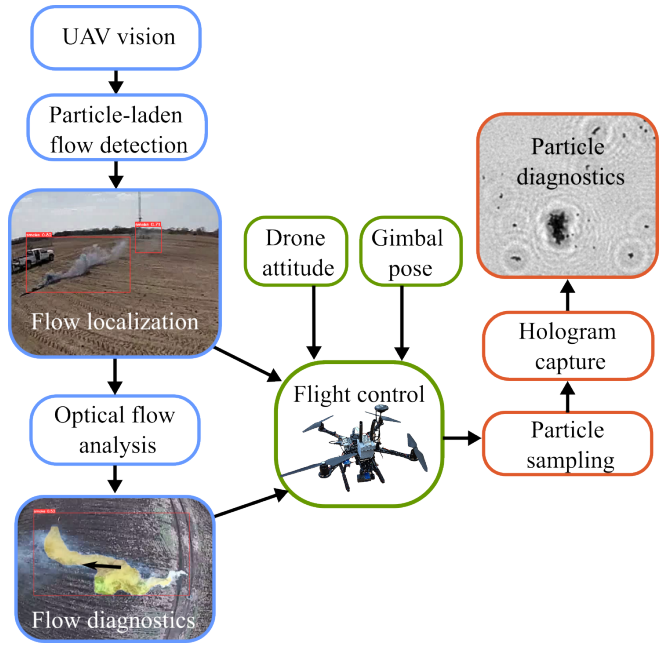

Figure 1. *Autonomous UAV operation workflow for enabling vision-based navigation and holographic imaging for in situ particle diagnostics.*

### A. Related work

In the case of wildfires, there have been significant collaborative research efforts in recent years with prescribed burn experiments (e.g., FASMEE, FIREX-AQ). These have generated valuable data concerning emissions and fire dynamics using a wide array of sensors, such as lidar, airborne gas sensors, satellite imaging, radiosondes, etc. [13]. Despite the level of control that is afforded by conducting prescribed burns (as opposed to measurements during natural wildfires), it is still not possible to characterize the size and shape of particulate matter on an individual particle basis across the broad spatial domain covered by the emissions, particularly at lower altitudes below which aircraft cannot sample. Remote

*Research supported by National Science Foundation.
N. R. Bristow is with the Mechanical Engineering department, University of Minnesota – Twin Cities, Minneapolis, MN 55414 (e-mail: nbristow@umn.edu).

N. Pardoe is with the Minnesota Robotics Institute, University of Minnesota – Twin Cities, Minneapolis, MN 55414 (e-mail: pardo020@umn.edu).

J. Hong is with the Mechanical Engineering department, University of Minnesota – Twin Cities, Minneapolis, MN 55414 (phone: 612-626-4562; e-mail: jhong@umn.edu).

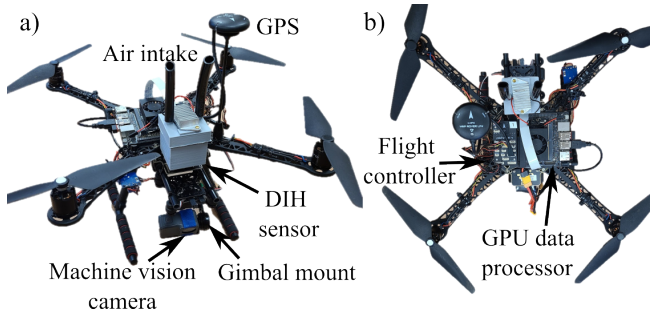

*Figure 2. UAV hardware including all on-board sensor, with a top-down view shown in (b).*

sensing with lidar can measure across $O(10^3$ m) scale domains, but coarse resolution misses spatial heterogeneities as the plume evolves.

Traditional PM sensors (e.g., Plantower PMS5003) can measure *in situ* but typically must make assumptions about individual particle morphology (e.g., spherical shape) [8, 14]. Other measurement approaches that may be able to provide greater detail, such as traditional microscopes, are too bulky and not amenable to being moved across a large measurement region. Alternatively, particles may be sampled in the field and then brought back to the lab [10], but this is also limited by the number of sampling locations and devices available, in addition to the drawback of not being able to measure *in situ*.

Digital inline holography (DIH) offers a promising alternative as a microscopic measurement technique for particle diagnostics for wildfires and other similar cases [15-18]. This imaging approach involves illuminating particles of interest (e.g., soot, dust, droplets) with a coherent light source, such as a laser. The light scattered off the particles constructively and destructively interferes with the coherent light source and results in light and dark fringe patterns to appear when captured by the camera sensor. These interference patterns can embed information concerning particle shape, size, and even composition. The in-focus particle image can be digitally "reconstructed" across a large depth of field compared to conventional techniques such as brightfield microscopy, making the optical arrangement more robust.

DIH has been implemented in aerial applications to measure droplets, dust, or pollen [11, 12], but usually with optical systems too bulky to be effectively integrated with an autonomous aerial vehicle (UAV). [12] demonstrated the use of DIH with a UAV, but in their system, the DIH particle sensor was large enough that it had to be carried far below the UAV on a tether. The sensor was also very heavy (~3 kg), such that it needed a large hexacopter UAV to carry the payload. Such implementations lack mobility and autonomy and thus are inherently limited in their ability to optimally sample particle properties as they disperse in the atmosphere over large distances.

Lensless imaging has been used to develop lightweight, compact DIH sensors [19, 20]. In such systems, the imaging resolution is limited primarily by the camera pixel size. Partially coherent light sources (e.g., an LED) can also be used instead of a laser, which results in better image quality with reduced speckle noise that otherwise appears in holograms due to the high coherence of laser light. The use of an LED requires only for the light to be passed through a pinhole, which, if the sample-to-sensor distance is small compared to the sensor-to-pinhole distance, can be relatively large (e.g., 100 μm). A large pinhole is particularly advantageous as it does not require careful alignment, a point that is important when considering applications in robotics with sensor vibration.

Lensless imaging with DIH has seen limited application in robotics. A notable recent work is by [21] who paired a laser-illuminated lensless DIH sensor with an aquapod for monitoring microparticles (e.g., algae cells, plastics, etc.) in water. However, their robot did not actively search for particles, and the data obtained by the robot's sensors, including the DIH sensor, was not used for feedback control. Beyond the fact that it was not designed for aerial applications, such a system lacks autonomy and, therefore, the ability to intelligently and optimally sample particle information as they spread and disperse.

## II. AUTONOMOUS UAV

### A. System overview

To achieve the aims of measuring atmospheric aerosol particulate matter with great detail and over a large domain, an autonomous UAV system was developed with the workflow depicted in Fig. 1. The measurement system is capable of searching for and then moving along with a particle-laden flow (e.g., a smoke plume) using autonomous vision-based navigation to enable optimal particle sampling. The target flow is first localized with object detection, wherein the output bounding box size and center can be used to guide the UAV toward the target. This is achieved by also integrating information available concerning the UAV attitude and pose of the camera gimbal. Once the UAV has completed its approach to the target, the flow direction and magnitude are determined using optical flow and background segmentation. Thus, with the flow localized and its trajectory determined, the UAV can immerse itself in the plume and move along with the particles, sampling and analyzing their properties *in situ* using digital inline holography. As such, the UAV can intelligently navigate itself to move along with the particles of interest such that changes in particle properties can be measured during the atmospheric dispersion process. In this manner, it is possible to capture individual microscopic particle details, $O(10^{-6}$ m), across the range of the UAV flight capabilities, $O(10^3$ m).

This measurement framework has been implemented on a Holybro S500 V2 quadcopter frame, as depicted in Fig. 2. The UAV is capable of carrying ~1 kg payload (after excluding the frame, motors and propellers, battery, flight controller, and GPS) for 15 minutes of flight with a 5200 mAh Lithium polymer battery. The payload of the current system is approximately 800 g, which includes the onboard GPU data processor, machine vision camera, gimbal mount, DIH sensor, and associated electronic components for delivering power. The GPU data processor used herein is an NVIDIA Jetson Xavier NX developer kit (hereafter referred to simply as the NVIDIA Jetson), which provides 384 NVIDIA CUDA Cores, 48 Tensor Cores, 6 Carmel ARM CPUs, and two NVIDIA Deep Learning Accelerators (NVDLA) engines. A GoPro Hero 7 Silver provides UAV vision and is mounted on a 3-axis Storm32 gimbal. The details of the DIH sensor are provided in Section III.

Flight is controlled using a Holybro Pixhawk 4 running ArduPilot and can receive commands both through a paired

2.4 GHz FrSky RC controller, 915 MHz telemetry radio via MAVLink, or directly over USB from the NVIDIA Jetson. The RC controller enables manual UAV control for simple movements in loitering mode, and the telemetry radio enables flight monitoring over Mission Planner flight planning software, run on a laptop on the ground. However, the primary advantage of this system, and hence its choice over an off-the-shelf UAV (e.g., from DJI), is the access provided by the Ardupilot software platform on the flight controller to be guided by in-house code run on the NVIDIA Jetson.

mix of the "low," "medium," and "high concentration smoke" image sets. These were combined with 656 images from experiments, augmented with grayscale, hue, blur, and noise

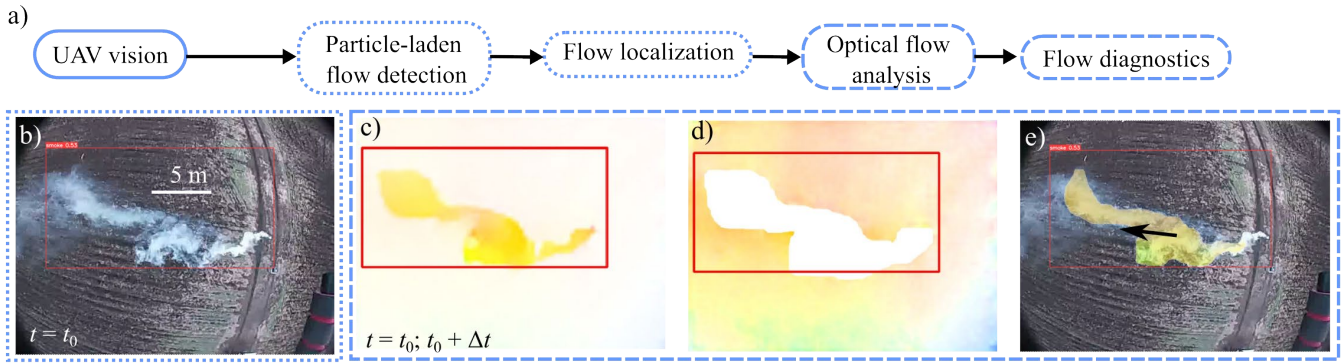

*Figure 3. Computer vision workflow for autonomous navigation of the UAV. (b) shows an example of object detection for smoke plume, and (c—e) show optical flow process, including segmentation.*

This feedback control for flying the UAV in guided mode is implemented with robot operating system (ROS) using the Python API. A series of nodes are concurrently run on the NVIDIA Jetson to enable vision-based navigation and flow diagnostics. MAVROS provides a MAVLink-enabled ROS node to access flight controller topics (e.g., setpoint velocity, GPS time, UAV attitude, etc.). ROS nodes that are run can subscribe or publish to flight controller topics as well as additional topics related to computer vision tasks. Thus, in addition to the MAVROS node, there are dedicated nodes for camera capture, object detection, optical flow, and general feedback control.

### B. Computer vision

The ROS node for UAV vision runs a continuous loop that streams 640 × 480 pixels$^2$ resolution RGB images at 30 fps over a wireless network connection between the GoPro and the NVIDIA Jetson via gopro-py-api [22]. Subscribing to GPS data via MAVROS, images are timestamped and geolocated, and the raw data is published to a ROS topic.

The object detection and optical flow nodes are depicted in Fig. 3. The detection node, marked with the dotted lines, subscribes to the most recently published image topic and uses a convolutional neural network with the "You only look once" (YOLO) architecture [23] in PyTorch (specifically, YOLOv5s) to localize a particle-laden flow, if present, from a single still image. In the current implementation, a smoke plume is used as the example case of a particle-laden flow. For this purpose, the object detection model was trained using synthetic and real-world smoke plume images.

As a foundation for the training dataset, we used a subset of the Smoke100k [24] synthetic images, which the authors therein generated using Blender to combine smoke patterns of various sizes and opacity overlaid on common background scenes. We further supplemented these with natural smoke images collected from prior experiments with the UAV. From Smoke100k, a total of 8000 images were taken, with an equal to provide 1968 total images from experiments. To improve the detection of darker smoke plumes, which were less common in the original dataset, both the synthetic and real-world experiment images were augmented by inverting 25% of the dataset. Approximately 15% of the final images did not contain a smoke plume to mitigate false positives. Training was performed using Google Colab for 50 epochs with a batch size of 32 to yield an mAP (Mean Average Precision) score of 99% with an IoU (intersection over union) threshold of 0.5. This PyTorch model was then converted to TensorRT on the NVIDIA Jetson for real-time detection at approximately 20-25 fps. The model's performance was found to be sufficient for the planned experiments involving real-world flight testing with smoke plumes created using smoke grenades, as detailed later. An example of this is shown in the first inset image of Fig. 3b.

Detection of a particle-laden flow in the image outputs a bounding box, the details of which are published as a ROS topic that the optical flow node subscribes to, along with the original image. Each time a new bounding box is published, the optical flow node waits for the next image topic, from which it computes the pixelwise movements in the entire frame. An example of this is shown in Fig 3c. Optical flow is computed via the Recurrent All-pairs Field Transforms (RAFT; [25]) algorithm, also implemented in PyTorch. We use a pre-trained model provided by the authors, which was fine-tuned on the Sintel dataset [14] and found to perform accurately enough for the smoke plume application. The model has an end-to-end speed of approximately one fps after post-processing [26].

The goal of performing optical flow is to determine the movement of the particle-laden flow and not of the background. Furthermore, it is desirable to determine the motion of the particle-laden flow independently of the apparent motion induced by the UAV or camera gimbal. We use k-means clustering to segment the optical flow results to accomplish both tasks. The bounding box information helps to disambiguate the foreground (i.e., particle-laden flow) from the background. The following approach was taken to compensate for the UAV or camera gimbal motion. First, the segmented particle-laden flow pixels were masked out (Fig. 3d), and the hole that was left was filled using a Navier-Stokes-based inpainting (i.e., interpolation) scheme in OpenCV that enforces a smoothness constraint. To minimize computational cost, the flow array was downscaled and then upscaled 4x before and after inpainting, respectively. The inpainted values were then subtracted from the originally computed flow values in a pixel-wise fashion to obtain a pixel-wise segmented motion for the particle-laden flow in a quasi-global framework, as depicted in Fig. 3e. Since this flow analysis was to be used to determine a setpoint velocity for the UAV, the segmented flow was averaged to obtain a single 2D output vector.

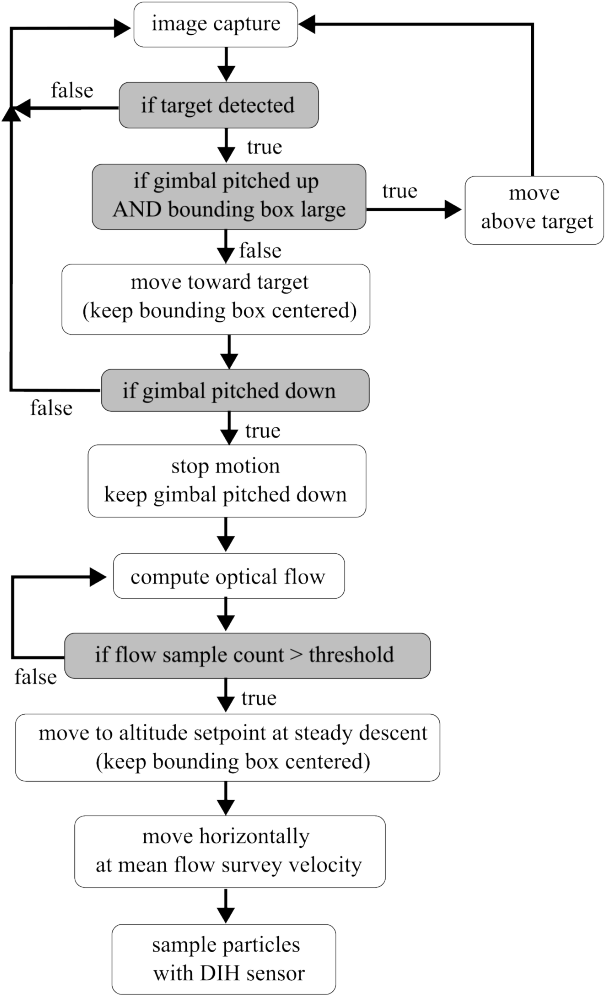

*Figure 4. Feedback control algorithm for approaching particle-laden flow, determining flow direction, and immersing UAV within particle-laden flow for optimal particle sampling.*

### C. Feedback control

With these key components in place, a specific feedback control loop was designed for the case of a smoke plume flow scenario (Fig. 4). In such a scenario, the UAV will search for smoke and initially approach it from the side. The approach is controlled by minimizing the bounding box offset from the image center and maximizing the bounding box size, using both linear motion and yaw control on the UAV, as well as gimbal pitch and yaw. The altitude is kept constant during this phase. As depicted in Fig. 4, if smoke is detected in the image, and the gimbal is pitched level (i.e., visibility is straight ahead), the UAV will continue to move directly towards the target. This continues until, as the UAV begins to pass over the top of the target, the gimbal will pitch downward in order to keep the target centered in its view. If the UAV is too low in altitude, however, and the bounding box achieves a size greater than 80% of either image dimension while the gimbal is still pitched level, the flight controller is told to raise its altitude by several meters and continue. This procedure loops until the UAV is at a high enough altitude to be able to approach the target from directly above. Once the gimbal is pitched downward beyond a threshold, the UAV is determined to be above the target and in an optimal position to perform flow diagnostics.

At this point, the UAV stops all motion, pitches the gimbal completely downward, and begins to analyze the smoke with optical flow. This "flow survey" period continues until sufficient samples of the particle-laden flow motion are collected to determine a meaningful average, as the smoke motion may fluctuate due to turbulence in the atmosphere. From this top-down view, the UAV begins a steady descent into the smoke plume towards a setpoint altitude such that the UAV, and the onboard DIH sensor, will be immersed in the smoke plume. During descent, the UAV's lateral motion is controlled with the bounding box information to ensure that it descends precisely into the center of the plume. If smoke is no longer visible in the image (e.g., due to UAV being too close to the smoke), the descent is maintained straight downward. Once immersed in the smoke at the setpoint altitude, the UAV begins to move laterally at the velocity determined by the optical flow survey previously conducted when above the smoke. The speed of motion is calibrated using the UAV altitude from the survey. As the UAV moves laterally within the smoke, PM in the plume are sampled using the DIH sensor, detailed in the following section.

### III. PARTICLE MEASUREMENT

Particle diagnostics are performed using digital inline holography (DIH) integrated with a pumped impaction system for sampling particles (Fig. 5a and b). Holographic imaging relies on coherent light interfering with light that has scattered off particles of interest, generating interference fringe patterns such as those shown in Fig. 5c. These holograms can be digitally reconstructed through convolution with the Rayleigh-Sommerfeld diffraction kernel to obtain in-focus images (Fig. 5d), with the advantage of a much larger possible depth of field than traditional imaging.

#### A. Hardware design

The system herein is specifically designed to obtain high-quality holograms with high throughput, capable of dealing with UAV-induced vibrations while minimizing payload. The hardware components include a 3D printed PLA enclosure, 3 W green LED and attached heatsink, a pinhole, a Raspberry Pi HQ camera, and a 150 L/min miniature air pump. The lensless Pi HQ camera features a Sony IMX477 sensor with a 1.55 μm pixel pitch and 4056 × 3040 pixel$^2$ resolution. The system operates by the pump, connected to the enclosure from beneath, drawing outside air into the chamber from two inlets at the top. These inlets are extended to draw in air from above the rotors using semi-rigid rubber tubing. The rotors on the UAV generate lift by pulling air downward toward the sensor. This air is then channeled into the enclosure by a pump. Inside the enclosure, an impaction nozzle focuses the flow of air and coherence to resolve $d = D\, z_1/z_2 = 1.55$ µm features, equal to the limiting pixel resolution. Furthermore, our use of a partially coherent light source also helps suppress coherence speckle noise, a common issue in laser-based holography caused by cross interference. The most significant advantage, however, is that the large size of the pinhole means that precise alignment between the LED and the pinhole is unnecessary. Therefore, the components are incredibly lightweight (~100 g), and the optics are not affected by UAV vibrations.

The design used herein increases particle throughput. There is no limitation of capturing particles due to camera shutter speed or exposure times. In fact, frame rates as low as desired are possible, and long exposure times can be used, which helps to compensate for the relatively low LED power (compared to a laser). Particles of size ranging between 3 µm and up to 50 µm could be detected by the sensor after impaction. The range of particle sizes captured by the impactor can be adjusted by varying the flow rate, and thus impaction speed, imposed by the pump. The imaging magnification can likewise be adjusted by varying the imaging objective to resolve particles ranging from 200 nm to a few millimeters in size.

### B. Image processing

Raw images are continuously captured during flight and georeferenced and timestamped to be synchronized with the machine vision camera. Processing of the holograms is currently performed offline but has the potential to be implemented using GPU acceleration for real-time analysis. Hologram reconstruction is performed after cropping images using a custom YOLOv5s model to provide bounding boxes around particle fringe patterns. The focused particle image, after reconstruction via the Rayleigh-Sommerfeld convolution, is segmented using k-means. The segmented particle image can then be used to compute various particle properties such as area, equivalent diameter, eccentricity, and roundness.

Another issue that must be overcome with the current DIH sensor is that particles accumulate on the sensor. To obtain clean images containing only new holograms at each time step, frame subtraction is used. At each pixel, the intensity from the previous image is removed, and then the grayscale value is adjusted.

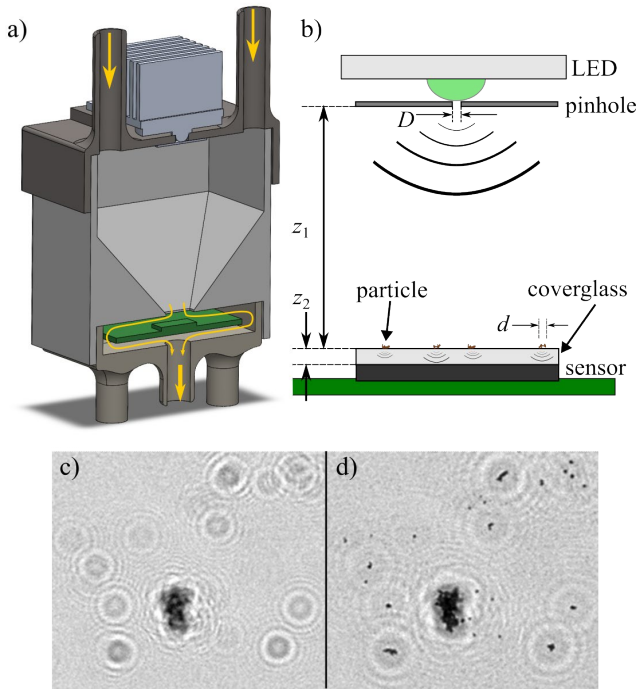

*Figure 6. Sensor for particle diagnostics using digital inline holography. (a) shows the internal structure of the enclosure, with airflow direction indicated with yellow arrows, where air is pumped in from the top. (b) depicts the optics of the lensless holographic imaging (not to scale). (c–d) shows a hologram examples obtained from the sensor in (a–b) before and after focusing the image with holographic reconstruction.*

directs it toward the camera sensor. A gap of only 1 mm between the nozzle and the sensor allows the air to divert sharply, causing any suspended particles, such as smoke soot, to be deposited onto the sensor's surface. Particles are illuminated with an LED that is butt-mounted to an ~100 µm diameter pinhole, as shown in Fig. 5b. As described by [27], a large $z_1/z_2$ ratio, where $z_1$ is the pinhole-to-sample distance and $z_2$ is the sample-to-sensor distance, results in an effective increase in the coherence of the LED light. More specifically, each particle can be treated as scattering incoming light that has passed through a pinhole, a factor of $z_1/z_2$ smaller than the physical pinhole. Thus, for our pinhole (diameter $D = 100$ µm), and for $z_1 = 45$ mm, $z_2 = 0.7$ mm, we have sufficient spatial

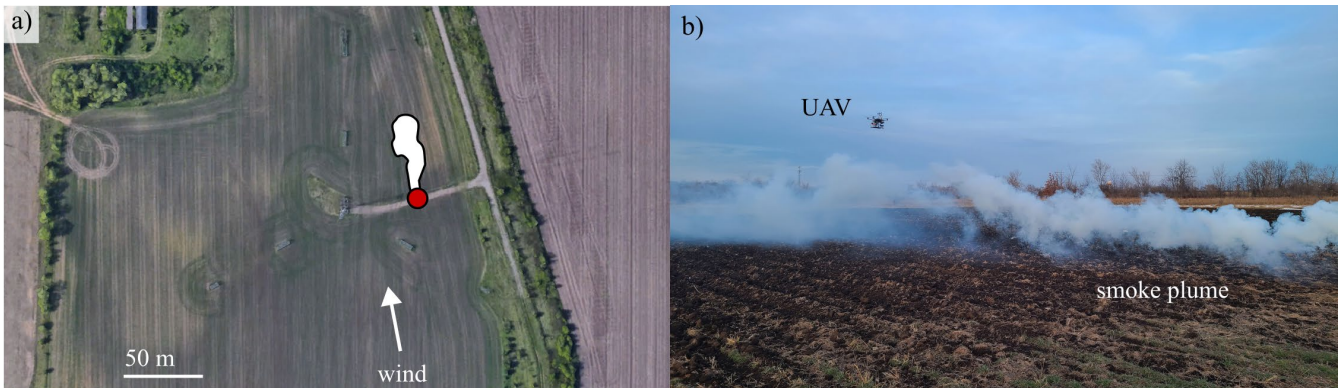

*Figure 5. (a) Map of experiment site at UMore Park in Rosemount, MN, with wind direction indicated with white arrow, red dot indicating smoke source, and white filled contour depicting scale of smoke plume. (b) Photo of an experiment with UAV in flight above a smoke plume from a grenade.*

## IV. EXPERIMENTS

The measurement system was tested at the University of Minnesota Outreach, Research, and Education (UMore) Park, on an agricultural field, as shown in Fig. 6. Smoke plumes were created with Enole Gaye EG18 high-output smoke grenades that were purchased in both white and black variants. Each grenade is capable of providing approximately 90 seconds of smoke emission, an example of which is shown in Fig. 6b. This duration is sufficient to test the capability of the autonomous flight mode, as the plumes created could grow up to 50-75 m long, depending on the wind speed, and the UAV's region of influence was observed to be confined within a few meters directly below the UAV.

Examples of the particle diagnostics possible with the UAV-based measurement system are shown in Fig. 7. In the dataset shown here, the UAV was flown manually back and forth within the plume. A 3D map of the sampling data can be seen in Fig. 7a, where colors indicate particle concentration by count, and the inset images display examples of holograms captured. These results show the expected trend that particle concentration is highest nearest the source and reduces as the particle-laden flow disperses downwind. Furthermore, the results shown in the inset images of Fig. 7a exemplify the level of detail afforded by DIH measurements, which is not possible with other PM sensors. Fig. 7b plots the same data as a function of distance from the source, with colors indicating the equivalent particle diameter (calculated as the $(4A/\pi)^{1/2}$, where $A$ is the area of the segmented particle image). These results demonstrate the heterogeneity in particle size that can be captured. Lastly, Fig. 7c shows the probability density functions of eccentricity and roundness for all particles captured.

Included in Supplemental Material is a demonstration of the autonomous flight operation. The video shows the images from the machine vision camera, overlaid with bounding boxes and later with segmented optical flow results during capture.

## V. CONCLUSIONS

By combining holographic imaging with UAV technology and machine learning-accelerated computer vision, we have demonstrated an intelligent measurement apparatus that can provide unique measurements not possible with any other system. While the demonstrations shown herein have involved smoke plumes, this is merely an example of the capabilities of autonomous UAV-based digital inline holography (DIH). The current system could also be easily modified to be applied to additional interesting cases, such as particulate matter from volcanic eruptions or dust clouds. Volcanic PM would be a particularly unique demonstration as the dangerous conditions surrounding a volcanic eruption necessitate the use of robots for *in situ* measurement, which has been used for monitoring gas emissions [28]. In the absence of UAV-based measurement, immobile ground stations or satellite measurements must be depended upon [29].

The ability of a DIH sensor to measure detailed particle characteristics raises the interesting possibility of also measuring microbes in smoke from natural wildfires. Recent work by [10] has shown that wildfires contain a substantial number of viable cells, compared to the ambient air, which they determine largely through laboratory processing of samples collected on a UAV. Using DIH, it is reasonable that a system like ours could analyze cells *in situ*, thus providing more precise georeferenced data.

An additional aspect of the system worth discussing is the potential influence of the UAV rotor wash on particle diagnostics. The flow induced by the quadcopter itself is capable of altering local flow conditions within a few meters, increasing turbulent mixing of the particle-laden flow. However, the scale of these motions induced by the UAV, $O(10^0\text{-}10^1$ m), is much smaller than the scale of the particle-laden flow itself, which for phenomena such as plumes from wildfires or volcanic eruptions is typically $O(10^3$ m), multiple orders of magnitude greater. Thus, it is unlikely for the UAV to notably influence the particle-laden flow at these large scales over which the UAV is going to be traversing and measuring trends. Instead, the downward laminar flow above the UAV rotors, where the air intake for the DIH sensor is located, is more likely to aid the sensor in capturing more particles.

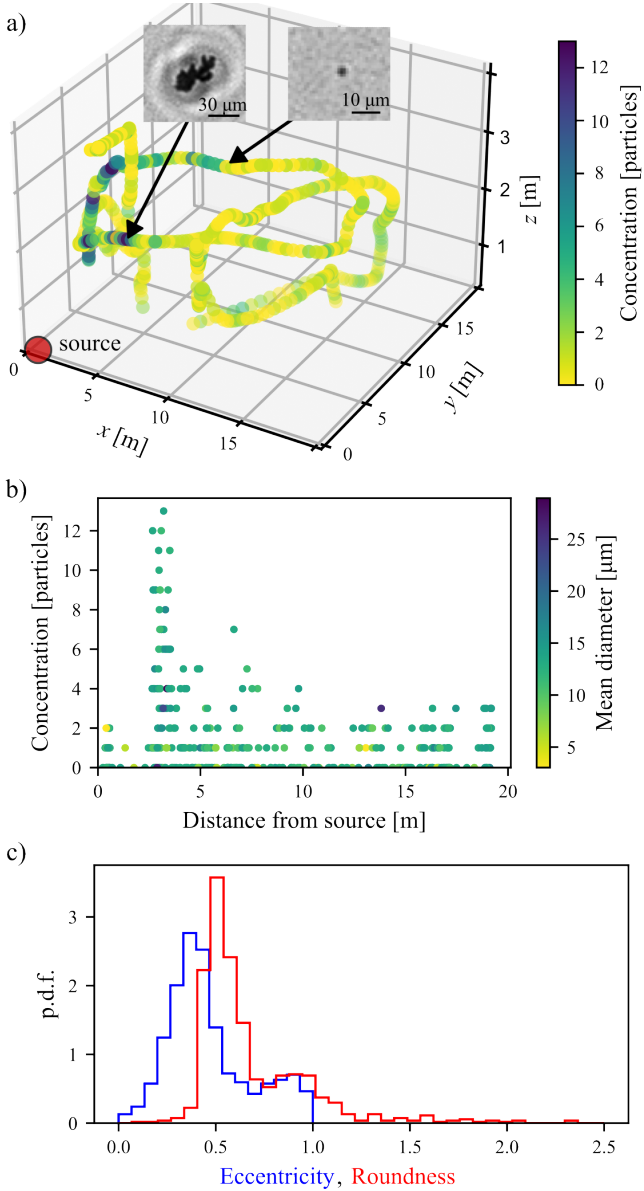

*Figure 7. Particle diagnostics data from the UAV-based DIH sensor. (a) spatial map of particle concentration along the UAV sampling path, with the source of smoke emission show in red and the inset plots depicting reconstructed smoke particle holograms. (b) concentration as a function of distance from the source, with marker color indicating mean equivalent diameter from each measurement location. (c) probability density functions of eccentricity and roundness for all particles sampled.*

It should also be noted that the same sensor can be used for particles other than the test case herein of smoke particulate matter, such as liquid droplets. The primary limitation is for larger sizes of particles, which may have too much inertia to be drawn into the sensor by the pump. However, given that the primary purpose of this measurement apparatus is for small particles or droplets suspended in the air (e.g., aerosols), this is not expected to be an important factor.

Although the present system is implemented as a single UAV operating on its own, it is possible to implement numerous UAVs with such sensors onboard in a swarm for more detailed and intelligent measurement capabilities. For example, upon determining the flow direction, a UAV swarm could align itself along the streamwise direction of the flow within the plume to instantaneously measure large-scale spatial gradients in particle characteristics rather than traversing across the entire flow. Furthermore, multiple UAVs could cover a larger domain than a single one could on its own.

The testing conducted herein was limited to relatively small smoke plumes and a single UAV. This system, and extensions of it using a UAV swarm, will be implemented in the future with large-scale prescribed burns to test the system's capabilities at full scale and demonstrate its capabilities to end users in the firefighting community who model and combat wildfires.

ACKNOWLEDGMENT

This work is supported by the National Science Foundation Major Research Instrumentation award CBET-2018658.